\documentclass{article}
\usepackage{graphicx}

\newcommand{\pd}{\partial}

\newcommand{\bdot}{\mbox{\boldmath $\cdot$}}
\newcommand{\del}{\mbox{\boldmath $\nabla$}}

\newcommand{\dv}{\mbox{\boldmath $\nabla \bdot$}}

\newcommand{\rh}{\hat{\rho}}
\newcommand{\etal}{et al.\ }

\begin{document}

\title{Magnetic Cycles and Meridional Circulation in Global Models of Solar Convection\footnote{To appear in
Proc.\ IAU Symposium 271, ``Astrophysical Dynamics, from Stars to Galaxies'', ed.\ A.S.\ Brun, N.H.\ Brummell
\& M.S.\ Miesch, in press}}

\author{Mark S. Miesch$^1$, Benjamin P. Brown$^2$, Matthew K. Browning$^3$, \\
Allan Sacha Brun$^4$ and Juri Toomre$^5$ \\
$^1$HAO, NCAR, Boulder, CO, 80307-3000, USA, email: {\tt miesch@ucar.edu} \\
$^2$Dept.\ of Astronomy, Univ.\ of Wisconsin,
475 N. Charter St., Madison, WI 53706, USA \\
$^3$CITA, Univ.\ of Toronto, Toronto, ON M5S3H8, Canada \\
$^4$DSM/IRFU/SAp, CEA-Saclay and UMR AIM, CEA-CNRS-Universit\'e Paris 7, \\
91191 Gif-sur-Yvette, France \\
$^5$JILA and Dept.\ of Astrophysical \& Planetary Sciences, 
Univ.\ of Colorado, \\ Boulder, CO 80309-0440, USA}

\maketitle

\begin{abstract}
We review recent insights into the dynamics of the solar convection
zone obtained from global numerical simulations, focusing on two 
recent developments in particular.  The first is quasi-cyclic magnetic
activity in a long-duration dynamo simulation.  Although mean fields
comprise only a few percent of the total magnetic energy they exhibit 
remarkable order, with multiple polarity reversals and systematic
variability on time scales of 6-15 years.  The second development
concerns the maintenance of the meridional circulation.  Recent
high-resolution simulations have captured the subtle nonlinear 
dynamical balances with more fidelity than previous, more laminar
models, yielding more coherent circulation patterns.  These patterns
are dominated by a single cell in each hemisphere, with poleward
and equatorward flow in the upper and lower convection zone respectively.
We briefly address the implications of and future of these modeling
efforts.
\end{abstract}
  
\section{Introduction}

As the Solar and Heliospheric Observatory (SOHO) was undergoing
its final stages of pre-launch preparations and as the telescopes 
of the Global Oscillations Network Group (GONG) were being deployed
around the world, Juri Toomre looked ahead with characteristic vision 
and enthusiasm:
\begin{quote}
The deductions that will be made in the near future from the 
helioseismic probing of the solar convection zone and the 
deeper interior are likely to provide a stimulus and to 
in turn be challenged by the major numerical turbulence
simulations now proceeding apace with the developments
in high performance computing (Toomre \& Brummell 1995).
\end{quote}
In the fifteen years since, the Michelson Doppler Imager (MDI) onboard
SOHO and the GONG network have provided profound insights into the
dynamics of the solar convection zone 
(Christensen-Dalsgaard 2002; Gizon \& Birch 2005; Howe 2009).
Meanwhile, high-resolution simulations of solar and stellar convection
have become indispensible tools in interpreting and guiding helioseismic
investigations 
(Miesch \& Toomre 2009; Nordlund \etal ~
2009; Rempel \etal ~ 2009).  Juri 
has played a leading role in both endeavors.

In particular, it was Juri who led a team of young students
and postdocs to develop what was to become known as the
ASH (Anelastic Spherical Harmonic) code
(Clune \etal ~ 1999; Miesch \etal ~ 2000). 
In last decade ASH has provided many novel insights into the
dynamics of solar and stellar interiors, including the
intricate structure of global-scale turbulent convection, the 
subtle nonlinear maintenance of differential rotation and meridional 
circulation, and the complex generation of mean and turbulent magnetic 
fields through hydromagnetic dynamo action
(Miesch \etal ~ 2000, 2006, 2008; Elliott \etal ~ 2000; Brun \& Toomre 2002; 
Brun \etal ~ 2004, 2005; Browning \etal ~ 2004, 2006; Browning 2008; Brown \etal ~
2008, 2010a,b; Brun \& Palacios 2009; Featherstone \etal ~ 2009).

In this paper we briefly review two recent development in ASH modeling,
focusing on the current Sun in particular.  These include the generation of 
coherent, organized, quasi-cyclic mean magnetic fields in an otherwise 
turbulent convection zone (\S2) and the delicate maintenance of the solar 
meridional circulation (\S3). Section 4 is a brief summary and look to the
future.  For ASH models of other stars and epochs, see the contributions 
by Brown and Browning (these proceedings).

\section{Magnetic Self-Organization in a Convective Dynamo}

Recent ASH simulations of convective dynamos in rapidly-rotating 
solar-like stars have revealed remarkable examples of magnetic 
self-organization, forming coherent toroidal bands of flux 
referred to as magnetic wreathes (Brown, these proceedings).  
There are typically two wreaths 
centered at latitudes of approximately $\pm 20-30$ degrees, 
with opposite polarity in the northern and southern hemispheres.   
They persist amid the intense turbulence of the convection zone,
maintained by the convectively-driven rotational shear.  Two
simulations in particular highlight the possible temporal 
evolution exhibited by such dynamos.  In Case D3, rotating
at three times the solar rate (rotation period $P_{rot} = $ 9.3 days), 
the wreathes persist indefinitely, buffeted by convective motions 
but otherwise stable for thousands of days (Brown \etal ~ 2010a).
In Case D5, rotating at five times the solar rate ($P_{rot} = $5.6 
days), the wreathes undergo quasi-cyclic polarity reversals on
a time scale of about 4 years (Brown \etal ~ 2010b).

\begin{figure}[t]
\begin{center}
 \includegraphics[width=\linewidth]{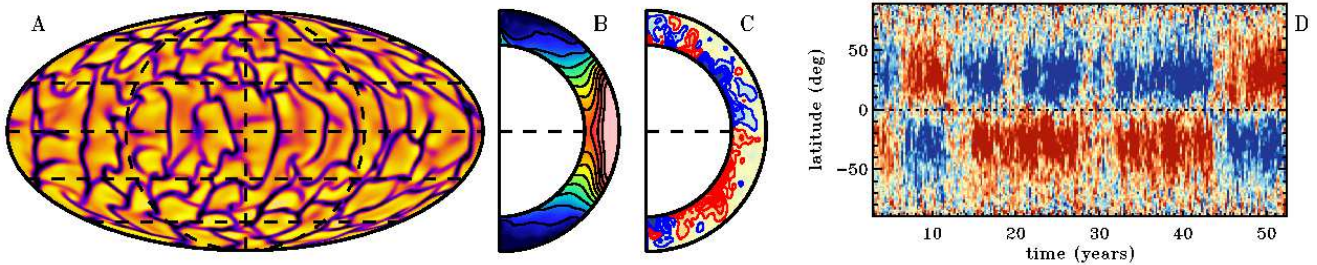} 
 \caption{Dynamo action in a solar convection simulation. (A) Radial velocity 
field near the top of the convection zone ($r = 0.96 R$) in a Molleweide 
projection (yellow denotes upflow, blue/black downflow).  (B) Angular velocity
$\Omega$ and (C) toroidal magnetic field $\left<B_\phi\right>$, averaged over 
longitude and time (pink/blue denote fast/slow rotation, red/blue denote 
eastward/westward field).  The time average for $\Omega$ spans the 50-year
time interval shown while the time average for $\left<B_\phi\right>$ 
focuses on a 6-day interval near $t = 25$ yr.
(D) Longitudinally-averaged toroidal field $\left<B_\phi\right>$ at the
base of the convection zone versus latitude and time.  Saturation levels
for the color tables are (A) $\pm$ 70 m s$^{-1}$, 
(B) 375--470 nHz, and (C,D) $\pm$ 3kG.  Peak values of $\left<B_\phi\right>$ 
reach $\pm$ 10--14 kG.}
\label{fig1}
\end{center}
\end{figure}

Such behavior is dramatically different from early ASH simulations 
of convective dynamos at the current solar rotation rate 
($P_{rot} = $ 28 days) by Brun \etal (2004).  These were 
dominated by small-scale turbulent magnetic fields with 
complex, chaotic mean fields.  The addition of a
tachocline promoted the generation of stronger, more
stable mean fields throughout the convection zone,
with wreath-like equatorially antisymmetric toroidal 
bands in the stably-stratified region below the convection zone 
(Browning \etal 2006; Miesch \etal 2009).
This tachocline simulation and a counterpart with different initial
conditions (Browning \etal 2007) were run for 
10-30 years each but no polarity reversals were observed.

Motivated by these previous studies, we initiated a new simulation 
in order to investigate whether self-organization processes comparable
to those exhibited by the rapid rotators D3 and D5 might occur also
at the solar rotation rate.  In order to enable long time integration,
we chose a moderate resolution of $N_r$, $N_\theta$, $N_\phi$ = 
129, 256, 512.  This is the same vertical resolution but half
the horizontal resolution of Case M3 from Brun \etal (2004).
The lower horizontal resolution required higher viscous,
thermal, and magnetic dissipation but by having the eddy dffusivity
scale as $\eta(r) \propto \rh(r)^{-1}$ rather than 
$\eta(r) \propto \rh(r)^{-1/2}$ as in case M3, we were able
to achieve comparable values of $\eta$ at the base of the
convection zone in the two cases; 
$\eta(r_b) = 2.03 \times 10^{11}$ cm$^2$ s$^{-1}$ in
Case M3 versus $\eta(r_b) = 2.88 \times 10^{11}$ cm$^2$ s$^{-1}$
in the case reported presently, which we refer to as Case M4.  
Here $\rh(r)$ is the background density profile and $r_b$ is the 
base of the convection zone.

Other than the horizontal resolution and dissipation, the principle
difference between M3 and M4 is the lower magnetic boundary condition.
Case M3 matched to a potential field whereas Case M4 employs a 
perfect conductor.  This has large implications for mean field 
generation; we find that perfectly conducting boundary conditions
promote the generation of strong toroidal fields in wreath-building
dynamos such as D3 and D5.  Furthermore, we have also imposed a
latitudinal entropy gradient at the lower boundary as described
by Miesch \etal (2006) in order to promote a solar-like, conical
angular velocity profile.  This takes into account thermal coupling
to the expected dynamical force balance in the tachocline without
explicitly including the tachocline itself, which requires fine
spatial and temporal resolution.  Omission of the tachocline may
well have important consequences as to the nature of the dynamo,
but we focus here on the generation of large-scale fields by
turbulent convection and rotational shear in the solar envelope
as an essential step toward understanding the fundamental elements
of the solar dynamo.

Results for case M4 are illustrated in Figure 1.  The convective
patterns (A) are similar to previous simulations of comparable 
resolution and the differential rotation (B) is solar-like,
with nearly conical mid-latitude contours and a monotonic
decrease in angular velocity of about 25\% from equator to
pole (475 nHz - 370 nHz).  In contrast to case M3, this 
simulation generates coherent layers of toroidal magnetic
flux near the base of the convection zone, with opposite
polarity in each hemisphere (C).  As the simulation evolves,
the polarity of these flux layers reverses several times (D).
The characteristic time scale for reversals appears to be 
about 14-15 years, although there is a failed reversal
at $t \approx 30$ yr.  The more rapid reversals early
in the simulations may represent initial transients as
the dynamo is becoming established.  The magnetic diffusion
time scale $\tau_\eta \sim r^2 \eta^{-1} \pi^{-2} $
ranges from 40 years at the inner boundary to 1.4 years 
at the outer boundary.

Given the generation of coherent mean fields, one might be 
tempted to refer to this as a large-scale dynamo 
(Brandenburg \& Subramanian 2005).  
However, the magnetic 
energy spectrum does not peak at large scales, as one may
expect from a turbulent $\alpha$-effect or an inverse cascade 
of magnetic helicity.  Rather, the magnetic energy peaks on
scales smaller than the velocity field, with mean fields making
up only 3\% of the total magnetic energy.  This is in contrast 
to the wreath-building rapid rotators D3 and D5 where 
approximately half (46-47\%) of the magnetic energy is 
in the mean fields.

In cases D3 and D5 the wreathes form and persist in the midst
of the turbulent convection zone whereas in case M4 the 
coherent toroidal fields are confined to the base of the
convection zone.  This can be attributed largely to the relative
strength of the rotational shear.  In cases D3 and D5, the
bulk of the kinetic energy (relative to the rotating frame) 
is in the differential rotation (65\% and 71\% respectively).
The stronger shear in turn is a consequence of the stronger 
rotational influence, which promotes angular momentum transport 
by means of the Coriolis-induced Reynolds stress.   In case M4 the rate 
at which the wreathes are generated by rotational shear is comparable
to or larger than the rate at which they are destroyed by
convective mixing.  Here only
35\% of the kinetic energy is in the differential
rotation, with 65\% in the convection (the meridional circulation
accounts for less than 1\% of the KE in all three cases).
The rotational shear is not strong enough to sustain the
wreathes in the mid convection zone and horizontal flux is pumped 
downward by turbulent convective plumes.  Toroidal flux accumulates
and persists near the base of the convection zone where the 
vertical velocity drops to zero and where it is further amplified 
by rotational shear.  The depth dependence of $\eta$ also 
contributes to the localization of the wreathes near the base
of the convection zone.  As mentioned above, $\eta \propto \rh^{-1}$
in case M4 whereas $\eta \propto \rh^{-1/2}$ in cases 
D3 and D5.

It is notable that such toroidal flux layers develop even
without a tachocline.  In the penetrative simulations by 
Browning \etal~(2006) radial shear in the tachocline does
contribute to the formation of persistent toroidal flux
layers but case M4 demonstrates that latitudinal shear
in the lower convection zone is sufficient.  This
is consistent with recent mean-field dynamo models
which suggest that latitudinal shear in the lower
convection zone is more effective at generating 
strong, latitudinally-extended toroidal flux layers
than the radial shear in the tachocline (e.g.\ Dikpati
\& Gilman 2006).  

Latitudinal shear is sufficient to generate toroidal flux layers at
the base of the convection zone but are they strong enough to spawn
the buoyant flux structures responsible for photospheric active
regions?  The peak strength of the mean toroidal field in case M4
reaches about 10$^4$G.  This is toward the lower end of estimated
field strengths of 10$^4$-10$^5$G based on observations of bipolar
active regions coupled with theoretical and numerical models of flux
emergence (Fan 2004; Jouve \& Brun 2009).
However, local (pointwise) values of the longitudinal field $B_\phi$
typically reach 30-40kG in the magnetic layers and even stronger
fields would be expected in higher-resolution simulations with less
subgrid-scale diffusion.  Convection may also promote the
destabilization and rise of flux tubes, producing solar-like tilt
angles of bipolar active regions even for relatively weak fields or
order 15 kG (Weber \etal ~ 2010).  Case M4 does not
exhibit buoyant flux structures, possibly due to insufficient
resolution.

A prominent feature of the rapidly-rotating wreath-building
dynamos D3 and D5 is an octupolar structure for the mean
poloidal field (Brown 2010a,b).  Poloidal separatrices lie 
at the poleward edge of the wreathes, although the magnetic
topology can become more complex during reversals.
By contrast, the mean poloidal field in case M4 is 
predominantly dipolar, often with weak, transient loops 
of opposite polarity at high latitudes.

We do not propose that this is a viable model of the solar dynamo.
It does not exhibit equatorward-propagating activity bands or
flux emergence comparable to active regions.  Still, it is remarkable
that the characteristic time scale for evolution of the mean fields
is just over a decade, two orders of magnitude longer than the
rotation period and the convection turnover time scale (both of
order a month).  Furthermore, quasi-cyclic polarity reversals
occur on decadal time scales without flux emergence (required 
by the Babcock-Leighton mechanism), without a tachocline, and 
without significant flux transport by the meridional circulation.
Cyclic variability on similar time scales was also found by
Ghizaru \etal (2010) in convective dynamo 
simulations with a tachocline.

\section{Maintenance of Meridional Circulation}

With the recent surge in popularity of Flux-Transport solar dynamo
models, the meridional circulation in the solar envelope has become
a topic of great interest.  According to the Flux-Transport paradigm,
equatorward advection of torodial flux by the meridional circulation
near the base of the solar convection zone is largely responsible
for the observed butterfly diagram and thus regulates the time scale 
for the 22-year solar activity cycle (Dikpati \& Gilman 2006;
Rempel 2006; Jouve \& Brun 2007; Charbonneau 2010).
In the postulated advection-dominated regime under which many
Flux-Transport models operate, the meridional circulation also
dominates the transport of magnetic flux from the poloidal source
region in the upper convection zone to the toroidal source
region near the base of the convection zone.

Kinematic mean-field dynamo models cannot address how the solar 
meridional circulation is maintained; it is imposed as a model 
input.  Non-kinematic mean-field models 
(e.g.\ Rempel 2005,2006) can give much insight 
into the underlying dynamics but they are necessarily based on 
theoretical models for the convective Reynolds stress, heat flux,
and $\alpha$-effect that need to be verified.  3D Convection simulations
are essential for a deeper understanding of how the solar meridional
circulation is established and what its structure and evolution
may be.

\begin{figure}[th!]
\begin{center}
 \includegraphics[width=\linewidth]{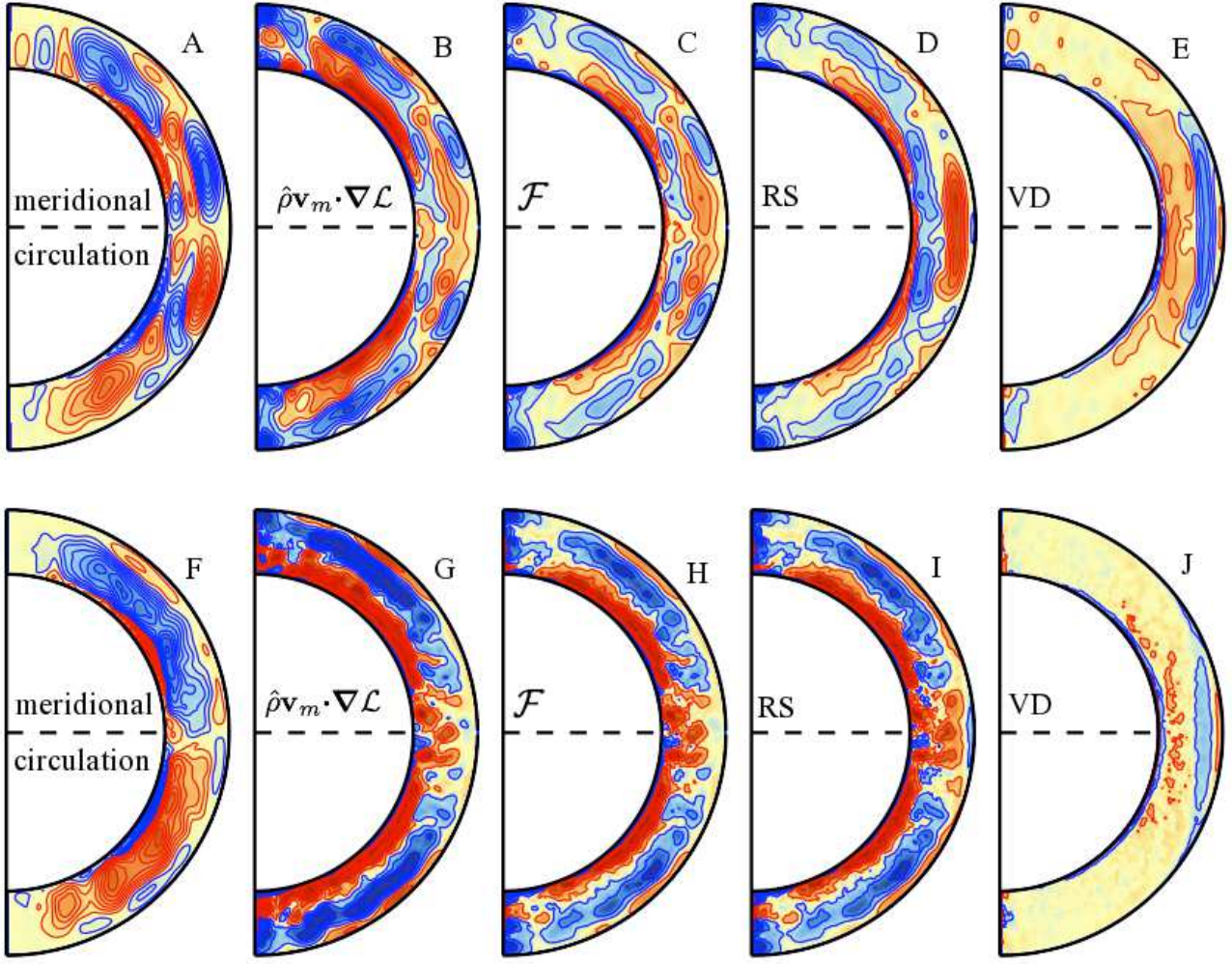} 
 \caption{Maintenance of meridional circulation in two solar
convection simulations.  The top row (A--E) shows the dynamo
simulation M4 discussed in \S2 while the bottom row (F--J) shows
a higher-resolution hydrodynamic simulation. Plotted are streamlines
of the mass flux (A, F), the left (B,G) and right (C,H) hand sides
of equation (\ref{gp}), and the contributions of the Reynolds stress
(D,I) and the viscous diffusion (E,J) to the net torque ${\cal F}$,
as expressed in eq.\ (\ref{torque}).  All quantities are averaged 
over longitude and time (each over 300 days) and contour levels 
range from $\pm 2\times 10^7$ g cm$^{-1}$ s$^{-2}$ in (B--E) and 
$\pm 5\times 10^6$ g cm$^{-1}$ s$^{-2}$ in (G--J). 
Contour levels for the streamfunctions in (A) and (F)
each range from $\pm 8\times 10^{21}$ g s$^{-1}$, corresponding to 
typical meridional flow speeds of order 20 m s$^{-1}$ and
several m s$^{-1}$ in the upper and lower convection zone
respectively.  Red and blue denote positive and negative values,
which correspond to clockwise and counter-clockwise circulations
in (A) and (F).}
\label{fig2}
\end{center}
\end{figure}

The meridional circulation in the solar envelope is established
in response to the convective angular momentum transport as follows
\begin{equation}\label{gp}
\left<\rh {\bf v}_m\right>_{\phi,t} \bdot \del {\cal L} = {\cal F} ~~~,
\end{equation}
where $\rh = \left<\rho\right>_{\phi,t}$ is the mean density stratification,
${\bf v}_m$ is the velocity in the meridional plane, and ${\cal L}$ is
the specific angular momentum:
\begin{equation}\label{amom}
{\cal L} = \lambda^2 \Omega = r \sin\theta \left(r \sin\theta \Omega_0
+ \left<v_\phi\right>_{\phi,t}\right)  ~~~.
\end{equation}
Angular brackets with subscripts $<>_{\phi,t}$ denote averages over 
longitude and time, $\Omega_0$ is the rotation rate of the rotating
coordinate system. $\Omega$ is the net rotation rate (including 
differential rotation), and $\lambda = r \sin\theta$ is the moment arm.
The net torque ${\cal F}$ includes components arising from the
convective Reynolds stress (RS), viscous diffusion (VD), and
the Lorentz Force:
\begin{equation}\label{torque}
{\cal F} = - \dv \left({\bf F}^{RS} + {\bf F}^{VD} + {\bf F}^{LF}\right) ~~~.
\end{equation}
where
\begin{equation}
{\bf F}^{RS} = \rh \lambda \left<{\bf v}_m^\prime v_\phi^\prime\right>_{\phi,t} ~~~,
\mbox{\hspace{.1in}}
{\bf F}^{VD} = - \rh \lambda^2 \nu \del \Omega
\mbox{\hspace{.1in}, and \hspace{.1in}}
{\bf F}^{LF} = - \frac{\lambda}{4\pi} \left<{\bf B}_m B_\phi\right>_{\phi,t} ~~~. 
\end{equation}
Here $\nu$ is the kinematic viscosity, ${\bf B}_m$ is the magnetic field
in the meridional plane, and primes denote variations about
the mean, e.g. $v_\phi^\prime = v_\phi - \left<v_\phi\right>_{\phi,t}$.

Note that equation (\ref{gp}) is derived from the {\em zonal} component of the
momentum equation yet largely determines the {\em meridional} flow.  
A convergence (divergence) of angular momentum flux, yielding a positive
(negative) torque ${\cal F}$, induces a meridional flow across ${\cal L}$ 
isosurfaces directed toward (away from) the rotation axis.  This is the concept 
of gyroscopic pumping; for further discussion and references see 
Miesch \& Toomre (2009).

The principle component of ${\cal F}$ responsible for maintaining the solar 
differential rotation and thus the meridional circulation via gyroscopic pumping 
is that due to the convective Reynolds stress ${\bf F}^{RS}$.  The viscous
component ${\bf F}^{VD}$ is negligible in the Sun.  However, in numerical
simulations viscous diffusion can largely oppose the convective Reynolds 
stress.  The meridional circulation then responds only to the residual 
torque, adversely influencing its structure and evolution.  This also
holds in the presence of the Lorentz force.  High resolution is required 
in order to minimize the artificial viscous diffusion and to therby achieve 
a more realistic zonal force balance.

The sensitivity of the meridional circulation profile to the zonal force 
balance is illustrated in Fig.\ \ref{fig2} for two ASH convection 
simulations.  The first (top row) is the moderate-resolution
dynamo simulation M4 discussed in \S2 ($N_r$, $N_\theta$, $N_\phi$ = 
129, 256, 512).  The second (bottom row) is a higher-resolution 
non-magnetic case with a lower viscosity that we will refer to as Case H
($N_r$, $N_\theta$, $N_\phi$ = 257, 512, 1024).  The difference in 
$\nu$ is roughly a factor of four (mid convection zone values 
are $3.3\times 10^{12}$ cm$^2$ s$^{-1}$ for case M4 versus 
$0.80\times 10^{12}$ cm$^2$ s$^{-1}$ for case H, both
varying with depth as $\rh^{-1/2}$).  Furthermore, the thermal 
diffusivity in case H is a factor of 2.7 less than in case M4
(mid convection zone values are $3.2 \times 10^{12}$ cm$^2$ s$^{-1}$ 
and $8.7\times 10^{12}$ cm$^2$ s$^{-1}$, again varying as
$\rh^{-1/2}$).

Comparison of panel pairs $B$--$C$ and $G$--$H$ indicate that the
dynamical balance expressed by eq.\ (\ref{gp}) is approximately
satisfied.  In both cases, the net torque ${\cal F}$ is given
by summing up the contributions from the Reynolds stress and
the viscous diffusion such that panel C $\approx$ D + E and
panel H = I + J.  The Lorentz force also contributes to the net
torque in panel $C$ but its amplitude is small relative to the
two terms shown (D,E).

There are two notable differences between the simulations M4 and 
H with regard to the maintenance of the meridional circulation.
First, the nature of the Reynolds stress is different (D, I).  
Both simulations exhibit a convergence of ${\bf F}^{RS}$ at low
latitudes that maintains the rotational shear but simulation
H also exhbits a much more prominent inward angular momentum
flux that is tending to accelerate the rotation in the lower 
convection zone relative to the upper convection zone.
Furthermore, there is a radially outward angular momentum
flux at low latitudes in case M4 that decelerates the lower
convection zone.  This is largely absent from case H.
The second difference is the relative contribution of the 
viscous diffusion, which plays a smaller role in Case H
due mainly to the smaller diffusivity $\nu$.

The net result of these two differences is that Case H
exhibits a substantial positive (negative) torque at
mid-latitudes in the lower (upper) convection zone.
The dynamical balance achieved in eq.\ (\ref{gp})
then induces a coherent single-celled meridional 
circulation pattern with poleward flow in the upper
convection zone and equatorward flow in the lower
convection zone (F).  By contrast,
the radially outward Reynolds stress near the equator
in case M4 and the opposing influence of 
viscous diffusion produce a weaker, less coherent
net torque ${\cal F}$.  The circulation patterns 
are consequently more complex, with multiple cells
in radius and latitude (A).

The circulation pattern in Case H (panel F) is predominantly
single-celled but exhibits narrow counter-cells near the 
boundaries.  These are likely artifacts of the boundary 
conditions.  At the bottom boundary we impose a latitudinal
entropy gradient as discussed by Miesch \etal (2006) in order
to take into account thermal coupling with the tachocline.
We also impose stress-free boundaries so the total angular 
momentum in the shell is conserved.  Although both boundary
conditions are justified, they are incompatible with thermal
wind balance whereby baroclinic and Coriolis-induced torques
offset one another.  The imbalance
induces a clockwise (counter-clockwise) circulation in the 
northern (southern) hemisphere near the boundary.  Penetrative 
convection simulations generally exhibit an equatorward
circulation throughout the overshoot region, induced by
the turbulent alignment of downflow plumes with the rotation
axis (Miesch \etal~2000).

A more realistic upper boundary condition would include coupling to
solar surface convection which includes granulation, mesogranulation
and supergranulation.  In lieu of the large Rossby number (rotation
period relative to the convective turnover time), one may expect
such surface convection to efficiently mix angular momentum, exerting a
retrograde zonal torque (${\cal F} > 0$).  The presence of such a
retrograde torque is implied by the existence of the near-surface shear layer, a
subsurface increase in rotation rate ($\pd \Omega/\pd r < 0$) detected
in helioseismic inversions (Howe 2009).  A retrograde torque in the
solar surface layers would induce a poleward circulation according to
eq.\ (\ref{gp}).  Thus, the upper and lower
counter-cells in the circulation profile of Fig.\ \ref{fig2}$F$ are
likely artificial.

\section{Summary and Outlook}

Inspired by helioseismology and fueled by continuing advances in
high-performance computing, global convection simulations 
continue to shape our understanding of solar internal dynamics and
the solar dynamo.  Recent simulations have achieved quasi-cyclic
magnetic activity on decadal time scales and meridional circulation
profiles that are qualitatively consistent with the single-celled
circulation profiles assumed in many kinematic mean-field dynamo models.
A more thorough discussion of these simulations and their implications 
will appear in forthcoming papers.

Further progress is imminent. The next decade may see the first global
convective dynamo simulations that spontaneously generate buoyant
magnetic flux structures from rotational shear that is self-consistently
maintained by the convection itself, providing unprecedented insights
into the origins of solar magnetic activity.  Achieving this milestone
will require numerical algorithms capable of exploiting next-generation
computing architectures with 10$^5$-10$^6$ processing cores.  It is a
daunting but glorious challenge; one that Juri will surely relish.

This research is supported by NASA through Heliophysics Theory Program
grants NNG05G124G and NNX08AI57G, and NASA SR\&T grant NNH09AK14I,
with additional support for Brown through NSF Astronomy and
Astrophysics postdoctoral fellowship AST 09-02004.  Browning was
supported by CITA and Brun was partly supported by the Programme
National Soleil-Terre of CNRS/INSU (France), and by the STARS2 grant
from the European Research Council. The simulations were carried out
with NSF PACI support of PSC, SDSC, TACC and NCSA, and by NASA HEC
support at the NASA Advanced Supercomputing Division (NAS) facility at
NASA Ames Research Center.

\appendix
\section{Discussion}

\noindent
ROGERS: I don't think counter-rotating cells at bottom
boundary are entirely artificial because of hard boundaries, we see
them in simulations with stable regions as well.

\vspace{.1in}
\noindent
MIESCH: The sense of the meridional flow near the lower 
boundary arises from the stress-free mechanical boundary condition
coupled to the imposed latitudinal entropy variation.  Our 3D penetrative
convection simulations generally have equatorward meridional flow in the
overshoot region as a result of the turbulent alignment of downflow 
plumes and gyroscopic pumping associated with the convective angular
momentum transport (Miesch \etal~ 2000).  2D simulations
may exhibit different behavior.  For further discussion, see the last
two paragraphs of \S3.

\vspace{.2in}
\noindent
HUGHES: Do you think all the physics for magnetic buoyancy
instability is incorporated in the anelastic approximation?

\vspace{.1in}
\noindent
MIESCH: The essential assumption behind the anelastic
approximation is that the Mach number is small.  In order for this to
break down in the deep solar interior within the context of flux
emergence, it would imply MG fields and order-one thermal
perturbations that can be ruled out by helioseismic structure
inversions.  The anelastic equation of state includes the influence of
the pressure on density variations so it captures the physical
mechanism underlying magnetic buoyancy.  For a derivation of the
Parker Instability in an anelastic system and nonlinear simulations
of rising flux tubes see Fan (2001).

\vspace{.2in}
\noindent
GOUGH: You commented that helioseismology indicates that the
concentrated polar vortex hich you illustrated as a property of some
your simulations poleward of about 85$^\prime$ does not exist. Actually that
is not so: helioseismology has nothing to say so close to the axis of
rotation. But it does indicate that the angular velocity poleward of
70° is lower than a smooth extrapolation would suggest- slow rotation
in the region where the large scale dipole like magnetic field
emanates and from which the fast component of the solar wind comes. Do
you find any indications of that in the simulations?

\vspace{.1in}
\noindent
MIESCH: The short answer is no.  Some of our simulations have
a monotonic decrease in the angular velocity $\Omega$ that continues
to the poles and some exhibit a slight increase in $\Omega$ toward the
poles.  In the former solutions, the slope is rather smooth, with no
indication of an abrupt steepening above 85$^\circ$.


\begin{thebibliography}{}

\bibitem{brand05}
{Brandenburg, A. \& Subramanian, K.}
2005, \textit{Phys.\ Rep.}, 417, 1

\bibitem{brown08}
{Brown, B.P., Browning, M.K., Brun, A.S., Miesch, M.S.\ \& Toomre, J.}
2008, \textit{ApJ}, 689, 1354

\bibitem{brown10a}
{Brown, B.P., Browning, M.K., Brun, A.S., Miesch, M.S.\ \& Toomre, J.}
2010, \textit{ApJ}, 711, 424

\bibitem{brown10b}
{Brown, B.P., Browning, M.K., Brun, A.S., Miesch, M.S.\ \& Toomre, J.}
2010, \textit{ApJ}, submitted

\bibitem{browning04}
{Browning, M.K., Brun, A.S.\ \& Toomre, J.}
2004, \textit{ApJ}, 601, 512

\bibitem{browning06}
{Browning, M.K., Miesch, M.S., Brun, A.S.\ \& Toomre, J.}
2006, \textit{ApJ Let.}, 648, L157

\bibitem{browning07}
{Browning, M.K., Brun, A.S., Miesch, M.S. \& Toomre, J.}
2007, \textit{Astron.\ Nachr.}, 328, 1100

\bibitem{browning08}
{Browning, M.K.}
2008, \textit{ApJ}, 676, 1262

\bibitem{brun02}
{Brun, A.S.\ \& Toomre, J.}
2002, \textit{ApJ}, 570, 865

\bibitem{brun04}
{Brun, A.S., Miesch, M.S.\ \& Toomre, J.}
2004, \textit{ApJ}, 614, 1073

\bibitem{brun05}
{Brun, A.S., Browning, M.K.\ \& Toomre, J.}
2005, \textit{ApJ}, 629, 885

\bibitem{brun09}
{Brun, A.S.\ \& Palacios, A.}
2009, \textit{ApJ}, 702, 1078

\bibitem{charb10}
{Charbonneau, P.} 2010,
\textit{LRSP}, 7, http://www.livingreviews.org/lrsp-2010-3  

\bibitem{clune99}
{Clune, T.L., Elliott, J.R., Miesch, M.S., Toomre, J., \& Glatzmaier, G.A.} 
1999, \textit{Parallel Computing}, 25, 361 

\bibitem{chris02}
{Christensen-Dalsgaard, J.} 2002, \textit{Rev.\ Mod.\ Phys.}, 74, 1073

\bibitem{dikpa06}
{Dikpati, M. \& Gilman, P.A}, 2006,
\textit{ApJ}, 649, 498

\bibitem{ellio00}
{Eliott, J.R., Miesch, M.S.\ \& Toomre, J.} 2000, 
\textit{ApJ}, 533, 546

\bibitem{fan01}
{Fan, Y.} 2001, \textit{ApJ}, 546, 509

\bibitem{fan04}
{Fan, Y.} 2004,
\textit{LRSP}, 1, http://www.livingreviews.org/lrsp-2004-1  

\bibitem{feath09}
{Featherstone, N.A., Browning, M.K., Brun, A.S.\ \& Toomre, J.} 2009, 
\textit{ApJ}, 705, 1000

\bibitem{ghiza10}
{Ghizaru, M., Charbonneau, P.\ \& Smolarkiewicz, P.K.\ 2010}, 
\textit{ApJ Let.}, 715, L133

\bibitem{gizon05}
{Gizon, L. \& Birch, A.C.} 2005,
\textit{LRSP}, 2, http://www.livingreviews.org/lrsp-2005-6  

\bibitem{howe09}
{Howe, R.} 2009, \textit{LRSP}, 
6, http://www.livingreviews.org/lrsp-2009-1  

\bibitem{jouve07}
{Jouve, L. \& Brun, A.S.} 2007,
\textit{A\&A}, 474, 239

\bibitem{jouve09}
{Jouve, L. \& Brun, A.S.} 2009,
\textit{ApJ}, 701, 1300

\bibitem{miesc00}
{Miesch, M.S., Elliott, J.R., Toomre, J., Clune, T.L., Glatzmaier, G.A.\ \&
Gilman, P.A.} 2000, \textit{ApJ}, 532, 593 

\bibitem{miesc06}
{Miesch, M.S., Brun, A.S.\ \& Toomre, J.} 2006, 
\textit{ApJ}, 641, 618 

\bibitem{miesc08}
{Miesch, M.S., Brun, A.S., DeRosa, M.L.\ \& Toomre, J.} 2008,
\textit{ApJ}, 673, 557

\bibitem{miesc09b}
{Miesch, M.S., Browning, M.K., Brun, A.S., Toomre, J. \& Brown, B.P.} 
2009, in: M. Dikpati, T.\ Arentoft, I.\ Gonz\'alez Hern\'andez, C.\ Lindsey
\& F. Hill (eds.),  \textit{Proc.\ GONG 2008/SOHO XXI Meeting on 
Solar-Stellar Dynamos as Revealed by Helio- and 
Asteroseismology}, ASP Conf.\ Ser., vol.\ 416, p.\ 443 

\bibitem{miesc09}
{Miesch, M.S., \& Toomre, J.} 2009,
\textit{Ann.\ Rev. Fluid Mech.}, 41, 317

\bibitem{nordl09}
{Nordlund, A., Stein, R.F.\ \& Asplund, M.} 2009,
\textit{LRSP}, 6, http://www.livingreviews.org/lrsp-2005-2

\bibitem{rempe05}
{Rempel, M. 2005},
\textit{Ap.J.}, 622, 1320

\bibitem{rempe06}
{Rempel, M. 2006},
\textit{Ap.J.}, 647, 662

\bibitem{rempe09}
{Rempel, M., Sch\"ussler, M., Cameron, R.H.\ \& Kn\"olker 2009},
\textit{Science}, 325, 171.

\bibitem{toomr95}
{Toomre, J., \& Brummell, N.H.} 1995, in: J.T.\ Hoeksema, 
V.\ Domingo, B.\ Fleck \& B.\ Battrick (eds.),
\textit{Fourth SOHO Workshop: Helioseismology} (ESA: Noordwijk), p.\,47

\bibitem{weber10}
{Weber, M, Fan, Y.\ \& Miesch, M.S.} 2010, in preparation

\end{thebibliography}
\end{document}